\documentclass[%
 reprint,
 amsmath,amssymb,
 longbibliography,
prb,
]{revtex4-1}

\usepackage{braket}
\usepackage{graphicx}
\usepackage{dcolumn}
\usepackage{bm}
\usepackage{xcolor}
\usepackage{mathtools}
\usepackage{amsmath}
\usepackage{ulem}
\usepackage{hyperref}
\hypersetup{
    colorlinks,%
    citecolor=blue,%
    linkcolor=blue,%
    urlcolor=blue
}

\newcommand{\fp}{{\it first-principles}}

\newcommand{\avg}[1]{{\langle #1 \rangle}}
\begin{document}

\title{Topological flat band, Dirac fermions and quantum spin Hall phase in 2D Archimedean lattices}

%
\author{F. Crasto de Lima}
\email{felipe.lima@ufu.br}
\author{Gerson J. Ferreira}
\author{R. H. Miwa}

\affiliation{Instituto de F\'isica, Universidade Federal de Uberl\^andia, \\
        C.P. 593, 38400-902, Uberl\^andia, MG,  Brazil}%

\date{\today}

\begin{abstract}
{Materials with designed properties arises in a synergy between theoretical and experimental approaches. In this study we explore the set of Archimedean lattices forming a guidance to its electronic properties and topological phases. Within these lattices, rich electronic structure emerge forming type-I and II Dirac fermions, topological flat bands and high-degeneracy points with linear and flat dispersions. Employing a tight-binding model, with spin-orbit coupling, we characterize a quantum spin Hall (QSH) phase in all Archimedean lattices. Our discussion is validated within density functional theory calculations, where we show the characteristic bands of the studied lattices arising in 2D carbon allotropes.}


\end{abstract}

\maketitle


\section{Introduction}

{Since the experimental discovery of graphene \cite{SCIENCENovoselov2004} two-dimensional (2D) materials have become the subject of numerous studies\cite{RSCALee2017, NATUREMannix2017}.}
{Correspondingly, the family of 2D materials have increased in recent years with the synthesis of new materials, e.g. transition metal-dichalcogenides \cite{NATUREManzeli2017}, borophene \cite{CSRZang2017}, gallenene \cite{SCIENCEKochate2018}, and many others\cite{AMGeng2018}. In synergy with the experiments, high throughput calculations have bring into evidence new 2D phases \cite{NATUREMounet2018,2DMATHaastrup2018}.}

{In these lower dimensional materials there is an emergence of phenomena prominent for applications. Particularly, the graphene Dirac fermions\cite{RMPCastro2009}, quantum spin, anomalous, valley and fractional Hall phases \cite{PhysRevLett.95.226801, SCIENCEHatsuda2018, SCIENCEKomatsu2018, PRBOstrovsky2008, PRLTang2011}, but also electronics based in the valley\cite{PRBAng2017}, layer\cite{JCPCrasto2019} and cone\cite{2DWu2016} degrees of freedom. On the other hand, monoatomicaly flat materials like graphene are still rare compared with the vast 2D materials class. Interesting properties arise in such class of materials, for instance, the perfectly flat structure of kagome and Lieb lattices presents a topological flat band\cite{PRBJiang2019} and fractional quantum Hall effect\cite{SRGong2014}. Moreover, high degeneracy points in the band structure allows exotic fermions to emerge\cite{SCIENCEBradlyn2016, JPCLWang2018, NPJPark2018, PRLOng2018}. Therefore, a focused research in new 2D lattices can uncover quasiparticles (e.g., of the Dirac, Weyl and Kane types) in condensed-matter physics.}

\begin{figure}[b!]
\includegraphics[width=0.9\columnwidth]{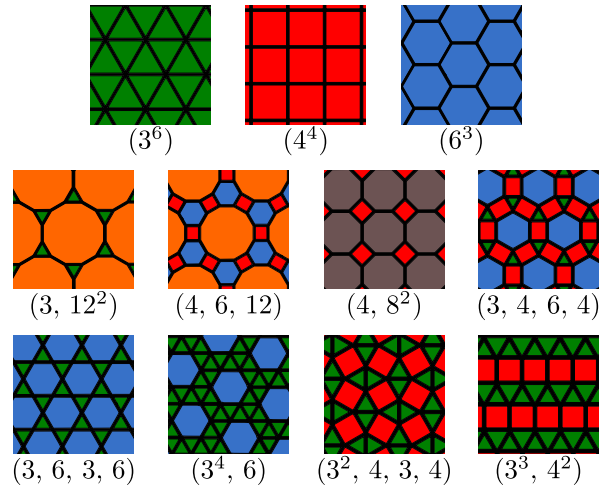}
\caption{\label{tillings} All eleven Archimedean lattices or uniform tilings, in which all polygons are regular and each vertex is surrounded by the same sequence of polygons.}
\end{figure}

{Although all possible periodical arrangement of sites in a plane is inconceivable large, there exist classes of 2D crystals that remain with its electronic properties unexploited. For instance, classes of lattices generated by considering one site in every vertex of the $k$-uniform tilling of the plane \cite{MATBranko1977}, where $k$ is the number of nonequivalent vertexes. The most simple lattices are given by $k=1$, i.e. only one nonequivalent vertex, namely uniform tiling, also known as Archimedean lattices, which are shown in Fig.\,\ref{tillings}. The notation taken to classify these lattice is to list the types of polygons forming each vertex and the times it appears, for instance (4,\,8$^2$) indicates that each vertex is surrounded by one square and two octagons. Particularly, the well known graphene (6$^3$) and kagome (3,\,6,\,3,\,6) lattices belong to the class of Archimedean lattices.}
		
{From a theoretical point of view Archimedean lattices have been explored showing magnetic frustration\cite{PREYu2015}, and possible formation in photonic crystals\cite{PRBUeda2007}, where its topological phases are still unexploited}. The design of photonic crystals allow highly controllable formation of structures, increasing the possible exploration of the Archimedean lattices\cite{NATUREKhanikaev2017}. { Additionally, beyond graphene and kagome lattices, there are other proposed and synthesized materials with Archimedean lattices, e.g. the graphenylene \cite{JMCCSong2013} with a (4,\,6,\,12) structure, and 2D allotropes of boron with (3$^4$,\,6)\cite{PRBTang2010, JPCLYi2017} and (3$^3$,\,4$^2$)\cite{JPCCLau2007} structures.} 
{Moreover, 2D metal-organic frameworks\cite{NATCHEMColson2013, CHEMCOMMUNRodriguez2016, CHEMREVStock2012} have recently been gaining attention. In such systems a particular lattice structure can be designed by combining different metals and molecular ligands.} Furthermore, a recent study has identified the distinct interparticle interactions needed to allow the self-assembly of the studied lattices\cite{PRLWhitelam2016}.

{In this paper we explore the electronic properties arising due to the flat structure of the Archimedean lattices. We provide a detailed band structure analysis with respect to single orbital tight-binding models. Therefore, our results can be directly interpreted in different quasiparticle systems, e.g. photonic \cite{NATUREKhanikaev2017}, phononic \cite{PRBXia2017, SCIREPYu2019}, magnonic \cite{PRBShindou2013}. Moreover, including the intrinsic spin-orbit coupling to the model, we explore the quantum spin Hall phases in electronic systems. Additionally, we validate our discussion presenting materials realizations of the Archimedean lattices.}

\section{Method}
 
To study the Archimedean lattices we consider a single orbital per site tight-binding (TB) Hamiltonian. Throughout the paper we will discuss first the systems without spin-orbit coupling (SOC), and subsequently in the presence of SOC\cite{PRLKane2005, PRLTang2011}. As the proposed systems are mirror symmetric in the plane of the sites, the Rashba SOC has null contributions. Therefore, the Hamiltonian can be written as
	\begin{equation}
	H_{TB} = H_0 + H_{SO}, \label{tb}
	\end{equation}
where each term is given by
\begin{equation}
H_0 = \sum_{i} \varepsilon_i c_{i}^{\dagger} \, \sigma_0 \, c_{i} + \sum_{i, j} t_{ij} c_{i}^{\dagger} \, \sigma_0 \, c_{j},
\end{equation}
\begin{equation}
H_{SO} = i \sum_{i, j} \lambda_{ij} c_{i}^{\dagger} \bm{\sigma} \cdot (\bm{d}_{kj} \times \bm{d}_{ik}) c_j.
\end{equation}
Here, $c_{i}^{\dagger} = (c_{i\uparrow}^{\dagger},\,c_{i\downarrow}^{\dagger})$ and $c_{i} = (c_{i\uparrow},\,c_{i\downarrow})^T$ are the creation and annihilation operators for an electron on site $i$ and spin $\uparrow$ or $\downarrow$; $\bm{\sigma} = (\sigma_x,\,\sigma_y,\,\sigma_z )$ are the spin Pauli matrices, $\bm{d}_{ij}$ is the vector connecting the $i$th and $j$th sites; $\varepsilon_i$, $t_{ij}$, $\lambda_{ij}$ are the on-site energies and strength of hopping and spin-orbit terms. In order to parametrize the Hamiltonian we have considered the hopping and spin-orbit terms exponentially decaying with the inter-site distance
\begin{equation}
t_{ij} = -N\,t \, \exp \left\{ - \alpha d_{ij} \right\},
\end{equation}
\begin{equation}
\lambda_{ij} = N\,\lambda \, \exp \left\{ - \alpha d_{ij} \right\}.
\end{equation}
All energies are normalized to the nearest-neighbor (NN) hopping, $ N = \exp \{ \alpha d_{\rm nn} \}$, with $d_{\rm nn}$ the NN distance, and taking $t=1$ as the energy unit. Within this definition the parameter $\alpha$ controls the contribution of the next nearest-neighbor (NNN) and, further neighbors, in relation to the NN. Therefore for $\alpha \gg d_{\rm nn}^{-1}$ only NN terms are relevant, while for small $\alpha$ the NNN and further terms strength increases.

\section{Results}

We will focus on the Archimedean lattices formed by more than one type of polygon tilling, as the first three lattices of Fig.\,\ref{tillings}, namely (3$^6$), (4$^4$) and the graphene (6$^3$) lattices are well known and extensively studied \cite{RMPCastro2009}. For the remaining eight lattices we have planar three fold coordination sites [(3,\,12$^2$), (4,\,6,\,12) and (4,\,8$^2$)], four fold [(3,\,4,\,6,\,4) and (3,\,6,\,3,\,6)] and five fold [(3$^4$,\,6), (3$^2$,\,4,\,3,\,4) and (3$^3$,\,4$^4$)]. Interestingly, all Archimedean lattices are composed by only one Wyckoff position\cite{Muller2004}. The hexagonal (3,\,12$^2$), (4,\,6,\,12), (3,\,4,\,6,\,4), (3,\,6,\,3,\,6) lattices transform as the $P6/mmm$ space group and are constructed by the Wyckoff point m, q, m and g respectively. The (3$^4$,\,6) hexagonal lattice transform as $P6/m$ space group and the k Wyckoff position. The square lattices (4,\,8$^2$) and (3$^2$,\,4,\,3,\,4) transform respectively as $P4/mmm$ and $P4/mbm$ space groups, with m and h Wyckoff positions. Lastly, the oblique (3$^3$,\,4$^2$) lattice transform as $Cmmm$ space group and with the j Wyckoff position.

As we will shown in the following sections, the Archimedean lattices present a rich electronic structure, ranging from flat bands to multiply degenerated Dirac cones. 
	
\subsection{Archimedean band structure}

In this section we discuss the Archimedean lattices band structure without SOC, which can be implemented in photonic crystals and direct related to other quasiparticle systems.

\begin{figure*}[t]
\includegraphics[width=2\columnwidth]{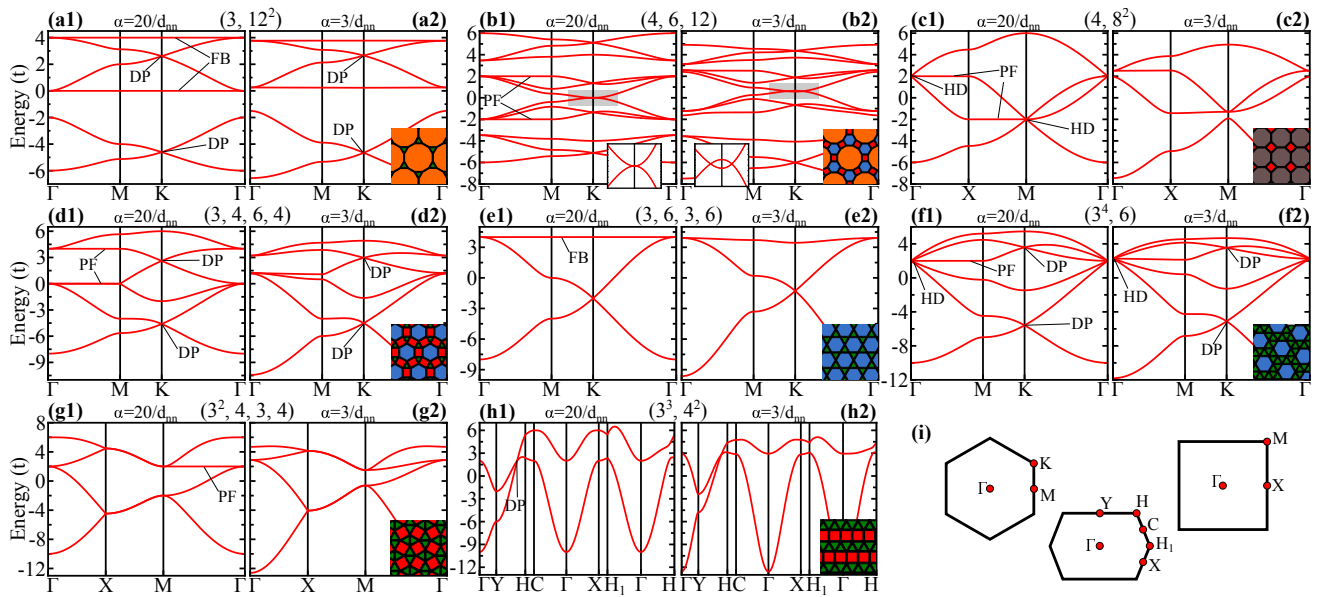}
\caption{\label{band-wo-soc} Band structure for $\alpha=20.0$/\,$d_{\rm nn}$ (a1)-(h1) and $\alpha = 3.0$\,$/d_{\rm nn}$ (a2)-(h2), without SOC ($\lambda = 0$). We also show the $1$st Brillouin Zones and the symmetry points (i) for the hexagonal, square and oblique lattices. Within the band plots we have labeled the Dirac points (DP), flat bands (FB), partialy flat (PF) bands and high degeneracy (HD) points.}
\end{figure*}

{Recently, the appearance of flat bands in materials was given a lot of attention, for instance in photonics\cite{APLPLeykam2018} due to its analogy with frustrated magnetism\cite{PRLLin2018} and possible fractional quantum Hall phase\cite{Sheng2011, PRBWang2012}. Interestingly, in the Archimedean lattices there are various occurrences of partially flat or fully flat bands. For the well known kagome (3,\,6,\,3,\,6) lattice\cite{JPCMBarreteau2017}, a perfectly flat band (FB) occur only for NN hopping [FB in Fig.\,\ref{band-wo-soc}(e1)], while considering further neighbor hoppings it gains a small dispersion [Fig.\,\ref{band-wo-soc}(e2)], increasing with the NNN hopping strength. A similar case can be observed in the (3,\,12$^2$) lattice, where two flat bands enclose dispersive graphene-like bands [FB in Fig.\,\ref{band-wo-soc}(a1)]. For this lattice the effect of NNN hopping change mostly the bandwidth without changing the flatness of the FB. 

Besides the perfect flat bands appearing in the whole Brillouin Zone, partially flat (PF) bands are also present in the studied systems. For instance, in the (4,\,6,\,12) lattice we can see flat bands occurring in the $\Gamma$-M direction in the 4th and, due to chiral symmetry, 9th band [PF in Fig.\,\ref{band-wo-soc}(b1)], while considering NNN hopping [Fig.\,\ref{band-wo-soc}(b2)] such symmetry is broken with the PF bands becoming dispersive. A similar case occur at the (4,\,8$^2$) and (3$^4$,\,6) lattices, where in the former the 2nd (3rd) band is flat in the X-M ($\Gamma$-X) direction [PF in Fig.\,\ref{band-wo-soc}(c1)], while in the latter the 4th band is flat in the $\Gamma$-M direction [PF in Fig.\,\ref{band-wo-soc}(f1)]. In this two lattices, considering further neighbors hopping do not present significant effects in the flat bands. The absence of the NNN effect on the flat bands is different for the (3,\,4,\,6,\,4) and (3$^2$,\,4,\,3,\,4) lattices, where the flat bands in the $\Gamma$-M [PF in Fig,\,\ref{band-wo-soc}(d1)] and M-$\Gamma$ [PF in Fig.\,\ref{band-wo-soc}(g1)] direction become dispersive for NNN hopping [Fig.\,\ref{band-wo-soc} (d2) and (g2)].}

{Another highly studied feature appearing in materials in general are the linear Dirac-like dispersions. Such energy spectrum is responsible for the high mobility in graphene, an many other phenomena\cite{RMPCastro2009}. In the class of Archimedean lattices, linear dispersive Dirac points (DP) are presented in different manners in all structures. For instance we can see two evident Dirac cone dispersions in the $K$-point of the lattices (3,\,12$^2$) [DP Fig.\,\ref{band-wo-soc}(a)], (3,\,4,\,6,\,4) [DP Fig.\,\ref{band-wo-soc}(d)] and (3$^4$,\,6) [DP Fig.\,\ref{band-wo-soc}(f)], which are still present even considering second neighbors hoppings. In the (4,\,6,\,12) lattice a symmetry analysis show that the $K$-point allow four doubly degenerated and two unidimensional irreducible representations. Looking at the NN band structure [Fig.\,\ref{band-wo-soc}(b1)] we find 5 points with double degenerated bands, which means that one of then is an accidental degeneracy. All except one (shaded region) of this double degenerated bands have linear cone-like dispersion. By taking NNN hopping into consideration we can see that the non-linear crossing opens a gap at the $K$-point [Fig.\,\ref{band-wo-soc}(b2) inset]. On the other hand, this two bands are still degenerated close to the $K$-point, forming a Dirac nodal line enclosing such point. 

Different from the symmetric type-I Dirac cone dispersion appearing in the kagome (3,\,6,\,3,\,6) lattice [Fig.\,\ref{band-wo-soc}(e1)-(e2)], in the (3$^3$,\,4$^2$) lattice we see a tilted linear crossing [DP in Fig.\,\ref{band-wo-soc}(h1)-(h2)], i.e. type-II Dirac quasiparticle. {Such feature, appearing also in borophene\cite{PRBPaul2019}, and transition-metal dichalcogenides\cite{PRLNoh2017, PRBHuang2016}, have a non-trivial Berry phase \cite{CMXu2018}}. Lastly, in the (3$^2$,\,4,\,3,\,4) lattice a linear, but not conical, dispersion can be observed along wave vectors crossing the X-M direction. Such feature form a nodal line in the border of the Brillouin Zone, with the linear dispersion for vectors crossing this line.}

{High degeneracy (HD) points in the band structure allows exotic fermions to emerge\cite{SCIENCEBradlyn2016, JPCLWang2018}. For instance, a hybrid state between a pseudospin $1$ and pseudospin $1/2$, also know as Kane fermion, have been observed in an 2D MOF \cite{JPCLWang2018}. A signature of such fermion is a flat band forming a three-fold degeneracy with a Dirac cone. Interesting we find in the (4,\,8$^2$) lattice two points with a pseudospin-1 Dirac dispersion, at the $\Gamma$ and $M$ [HD Fig.\,\ref{band-wo-soc}(c1)]. On the other hand taking the NNN hopping into consideration its degeneracy is lowered [Fig.\,\ref{band-wo-soc}(c2)]. A different scenario appears in the (3$^4$,\,6) lattice, where at the $\Gamma$ point we find two Dirac cones, with different velocities, degenerated with a flat band, forming a pseudospin-2 Dirac dispersion [HD in Fig.\,\ref{band-wo-soc}(f1)-(f2)]. In this case the degeneracy is still present even considering NNN hopping.} 

{In the following section we will discuss the effect of SOC to the Archimedean lattices band structures, and characterize the emergent quantum spin Hall phases.}

\subsection{QSH phase induced by SOC}

\begin{figure*}
\includegraphics[width=2\columnwidth]{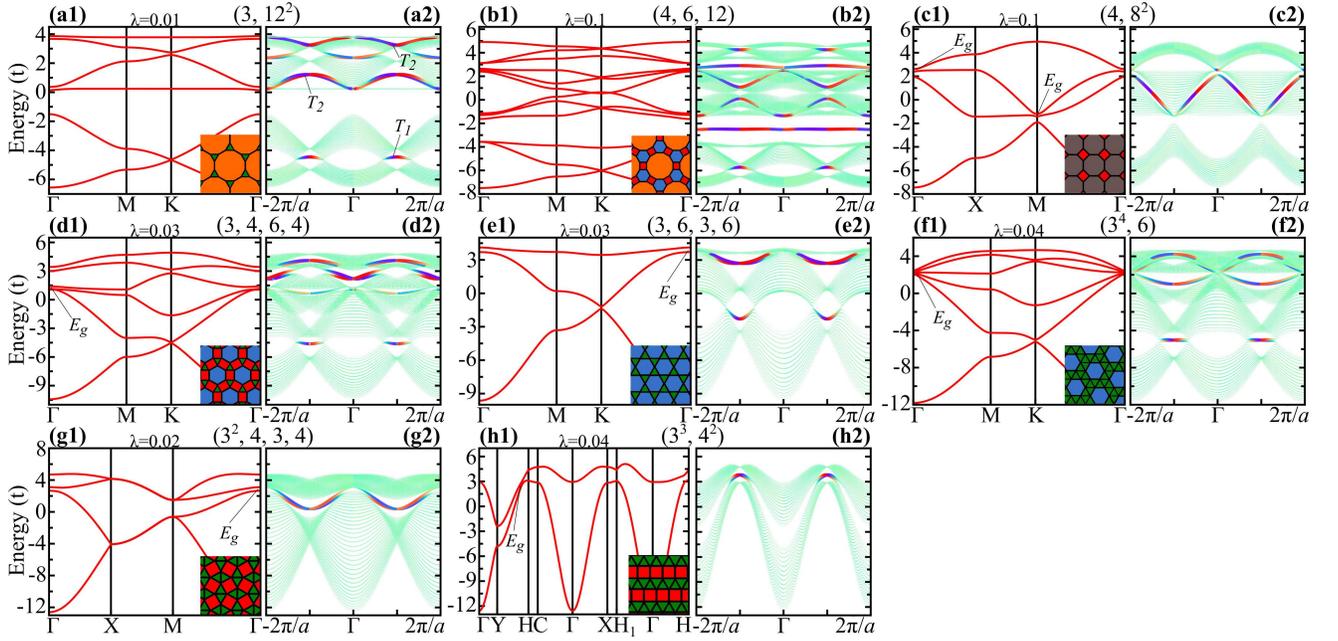}
\caption{\label{band-ribbon} Band structure with SOC and NNN hoppings ($\alpha = 3.0/d_{\rm nn}$) (a1)-(h1). Nanoribbons band structure with projected contribution of the edge sites (a2)-(h2), the size of each line is proportional to the edge contribution and the color is the $S_z$ spin projection. Within the plots, $E_g$ labels the energy gaps, while $T_1$ and $T_2$ the backscatering protected edge states.}
\end{figure*}

{In order to characterize the possible QSH phases in the Archimedean lattices, we take the SOC term into account. Usually, the opening of an energy gap, in previous degenerated bands, due to the atomic SOC, is a signature of QSH phase\cite{PRLKane2005}}.

By considering the SOC term, within the NNN hopping bands of Fig.\,\ref{band-wo-soc}(a2)-(h2) we see the gap opening on some of the degenerated points, as show in Fig.\,\ref{band-ribbon}. For instance, for all hexagonal lattices, at the Dirac points we see a gap opening. Additionally, the degeneracy points of a flat band and the Dirac cones at the $\Gamma$ point also presents a gap opening, e.g. the gap at $E=4.0$\,$t$ on the (3,\,6,\,3,\,6) lattice [$E_g$ in Fig.\,\ref{band-ribbon}(e1)]. In particular, the triple degeneracy of (3,\,4,\,6,\,4) $\Gamma$ point around $E=1.0$\,$t$ also breaks into three non-degenerated points [$E_g$ in Fig.\,\ref{band-ribbon}(d1)]. The same occur to the quintuple degeneracy at (3$^4$,\,6) $\Gamma$ point at $E= 2.5$\,$t$ [$E_g$ in Fig.\,\ref{band-ribbon}(f1)]. For the square lattice (4,\,8$^2$) the degeneracy of $\Gamma$ and $M$ points are broken, as in the non-symmorphic lattice (3$^3$,\,4,\,3,\,4) $\Gamma$ point [$E_g$ in Fig.\,\ref{band-ribbon}(c1) and (g1)]. In contrast, the degeneracy at $X$-$M$ direction arising from the non-symmorphic symmetry is preserved. Lastly the accidental degeneracy of the tilted Dirac cone in the oblique (3$^3$,\,4$^2$) lattice also become gaped with the inclusion of the SOC. 

\begin{figure}[h!]
\includegraphics[width=0.85\columnwidth]{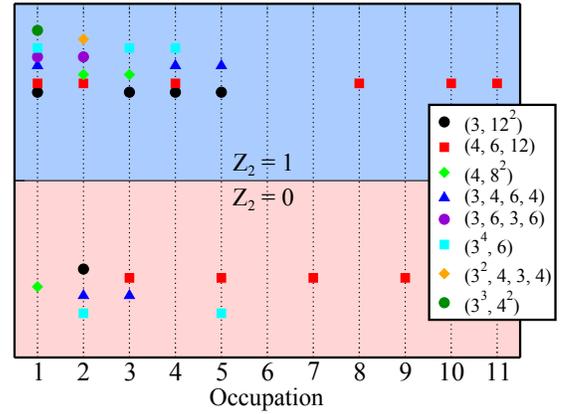}
\caption{\label{z2-occup} Calculated $Z_2$ invariant for different number of occupied  Kramers pairs bands. The occupation number indicates the number of occupied Kramer pairs bands starting in the lowest energy.}
\end{figure}
	
To characterize the existence of the QSH phases, we have tracked the Wannier charge centers evolution \cite{PRB84Bernevig, PRB83Vanderbilt} to determine the ${Z}_2$ topological invariant. A summary of the $Z_2$ index as a function of the Kramers pair bands occupation is shown in Fig.\,\ref{z2-occup}. Interestingly all studied lattices presents in some particular occupation a QSH phase. For instance, as stated, in all Dirac cones the SOC lead to a QSH edge state. To highlight a few cases, in the (3,\,12$^2$) lattice [black circles in Fig.\,\ref{z2-occup}] a $Z_2=1$ phase is observed for the occupation of the first Kramer pair of bands, i.e. the Fermi level being in the lower Dirac point of Fig.\,\ref{band-ribbon}(a1). Indeed, computing the edge states band projection we find spin-textured bands within the same energy window, $T_1$ in Fig.\,\ref{band-ribbon}(a2). Beside the Dirac points, the degeneracy point between a flat and dispersive bands also lead to the QSH phase, e.g. the 4th and 6th occupation of (3,\,12$^2$) lattice in Fig.\,\ref{band-ribbon}(a1) has $Z_2=1$ [Fig.\,\ref{z2-occup}]. In turn, it leads to the backscattering protected edge states, shown by $T_2$ in Fig.\,\ref{band-ribbon}(a2). A particular case occur in the hexagonal (4,\,6,\,12), where the SOC do not open the accidental degeneracy Dirac node between 6th and 7th band [Fig.\,\ref{band-ribbon}(b1)], which in turn do not have a defined $Z_2$ invariant. The same occur also for the nonsymmorphic symmetry protected degeneracy between $1$st and $2$nd bands of (3$^2$,\,4,\,3,\,4) [Fig.\,\ref{band-ribbon}(g1)].

The application of the results above rely on the synthesis of materials in Archimedean lattices structures. In the next section we discuss the material realization of the Archimedean lattices, while as a proof of principle we have characterized the existence of its characteristic band dispersions arising in carbon allotropes.

\subsection{Materials realization of Archimedean lattices}

It is interesting to point out that materials with Archimedean lattices are already proven theoretically and experimentally to exist. First, graphene\cite{SCIENCENovoselov2004}, the most studied 2D material, is the immediate example of the (6$^3$) Archimedean lattice. Experimentally, self-assembly of molecules in surfaces lead to a variety of possible structures, where structures equivalent to the (3,\,4,\,6,\,4)\cite{JMCCLyu2015}, (3$^4$,\,6)\cite{FDYan2017J} and (3$^2$,\,4,\,3,\,4)\cite{JPCCUrgel2014} have been observed. By using {\fp} calculations, it was predicted materials with the Archimedean lattices, e.g. graphenylene, an allotrope of carbon, presents the (4,\,6,\,12) lattice\cite{JMCCSong2013}. Also, a 2D boron sheet form the (3$^4$,\,6) structure\cite{JPCLYi2017}. Bellow, as a proof of principle, we present three other Archimedean lattices and show the presence of its characteristic electronic dispersion.

Here, by using {\fp} calculations, based in the density functional theory (DFT), we focus in the monoatomic layers of carbon allotropes. Within the VASP\cite{VASP} code we relax the atomic positions and lattice vectors, until the force in each atom was less than 1 meV/{\AA}. We have considered the PBE\cite{PBE} functional to describe exchange and correlation term, with the projector augmented-wave\cite{PAW} method for the electron-ion interaction. The wave functions were expanded in a plane-wave basis with cutoff energy of 500 eV. For BZ integration, a $11\times 11 \times 1$ k-mesh was considered. For the calculations of the force constants we considered a $3 \times 3$ supercell with energy convergence criteria of at least $10^{-8}$\,eV.

\begin{table}[h!]
\caption{\label{tab-carbon}Lattice parameter $a$ ({\AA}), average nearest neighbor distance $\avg{d_{cc}}$ ({\AA}), and its root mean-square deviation $\langle \delta d \rangle = \sqrt{\avg{d_{cc}^2} - \avg{d_{cc}}^2}$ ({\AA}). Formation energy $E_f = E_{\rm Arc} - E_{G}$ (eV/atom), where $E_{\rm Arc}$ is the total energy per atom of the Archimedean lattice and $E_G$ is the total energy per atom of graphene.}
\begin{ruledtabular}
\begin{tabular}{cccccc}
Lattice & state & $a$ & $\avg{d_{cc}}$ & $\avg{\delta d_{cc}}$ & E$_f$ \\
 \hline
(3,\,12$^2$) & flat & 5.143 & 1.401 & 0.033 & 0.969 \\ 
(4,\,6,\,12) & flat & 6.768 & 1.439 & 0.051 & 0.633 \\ 
(4,\,8$^2$)  & flat & 3.446 & 1.435 & 0.044 & 0.513 \\ 
\end{tabular}
\end{ruledtabular}
\end{table}

\begin{figure}[h]
\includegraphics[width=\columnwidth]{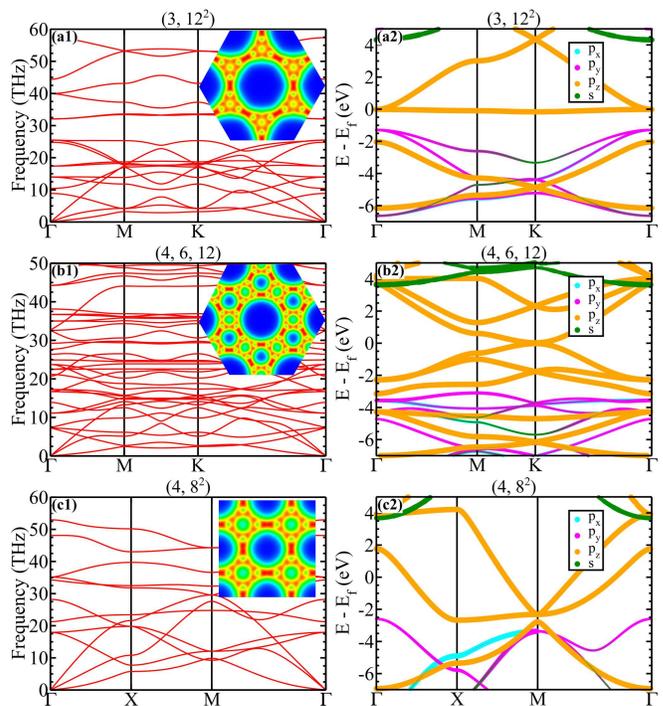}
\caption{\label{carbon} Phonon dispersion (a1)-(c1) and orbital projected band structure (a2)-(c2) for the three folded carbon lattices (a) (3,\,12$^2$), (b) (4,\,6,\,12) and (c) (4,\,8$^2$). The inset in the left side plot shown the charge density map in the plane with red (blue) indicate high (low) density regions.}
\end{figure}

In two-dimensions, carbon atoms usually form the planar 3-fold coordinated $sp^2$ hybridization. Naturally, satisfying this geometry, the carbon atoms could form the (3,\,12$^2$), (4,\,6,\,12) and (4,\,8$^2$) lattices. In fact, as shown in Tab.\,\ref{tab-carbon}, this carbon lattices present formation energy close to the already synthesized graphyne phase ($\sim0.7$\,eV/atom)\cite{CCLi2010, SRZhao2013}. To find the stability of the Archimedean carbon lattices we have calculated the phonon spectra, as shown in Fig.\,\ref{carbon}(a1)-(c1), where no negative frequencies were find, in agreement with previous study\cite{PCCPWang2012}. Furthermore, in the ground state structure we find mean carbon-carbon distance ($\avg{d_{cc}}$) with a deviation of only $~1\%$ in relation to the graphene lattice ($d_{cc}=1.42$\,{\AA}), as shown in Table\,\ref{tab-carbon}. However, a fluctuation in $d_{cc}$ is observed for nonequivalent bond directions, which we can see from the charge density map in the inset of Fig.\,\ref{carbon}(a1)-(c1), with dark red regions indicating greater charge density, correlated with smaller $d_{cc}$. Nevertheless, this bond distance variation do not modify the space group in relation to the perfect Archimedean lattice.

The planar geometry of this systems will lead to a decoupling of the carbon $p$ orbitals, in the even ($p_x/p_y$) and odd ($p_z$) in-plane mirror symmetries. Furthermore, the $p_z$ orbitals have azimuthal symmetry, therefore the interaction with neighbouring $p_z$ will be only distance dependent. The proposed TB hamiltonian of the second section has exactly this picture, in turn it will capture the bands formed by the carbon $p_z$ orbitals. First for the (3,\,12$^2$) lattice, note that the $p_z$ orbitals, orange circles in Fig.\,\ref{carbon}(a2), form the low energy Dirac crossing and structure of flat bands enclosing Dirac dispersion, found in Fig.\,\ref{band-wo-soc}(a1), with the flat band resonant in the Fermi energy. Such occupation configuration is exactly one that has a $Z_2=1$ phase with the inclusion of SOC. For the (4,\,6,\,12) lattice, Fig.\,\ref{carbon}(b2), the Fermi level is exactly in the accidental highlighted crossing shown in the inset of Fig.\,\ref{band-wo-soc}(b1). However, due to the higher bond distance variation ($\avg{\delta d_{cc}}$), this accidental degeneracy is lifted by a energy gap of $50$\,meV. Nonetheless, the Dirac crossings are still preserved in $2.3$\,eV  ($-1.8$\,eV) above (below) the Fermi level. Lastly, in the most stable (4,\,8$^2$) lattice, the $p_z$ orbitals show a breaking of the three-fold degeneracy in the $M$-point, Fig.\,\ref{carbon}(c2). Such feature indicates the relevance of NNN neighbours hopping, compare for instance with Fig.\,\ref{band-wo-soc}(c1)-(c2). Furthermore, external perturbations allow the control of these Archimedean bands. For instance, an isotropic strain field will change the atoms distance allowing a control of the NN and further neighbour hopping percentage, while proximity effects can induce SOC leading to the QSH effect\cite{NCAvsar2014}

\section{Summary} 
The Archimedean lattices presets rich electronic structure. For instance, we find type-I and II Dirac crossings and flat bands. Additionally, high degeneracy points, associated with pseudospin-1 and -2 Dirac quasiparticles have been found. The inclusion of spin-orbit couplings lead to topological energy gaps, where we construct a map to the occupation with backscaterring protected edge states. Finally, our lattice tight binding model is corroborated by showing, within {\fp} calculations, the existence of the characteristic Archimedean bands in stable 2D carbon allotropes. {These results work as a guide for exploration of designed 2D lattices.}

\acknowledgments
The authors acknowledge financial support from the Brazilian agencies CNPq, and FAPEMIG, and the CENAPAD-SP and Laborat{\'{o}}rio Nacional de Computa{\c{c}}{\~{a}}o Cient{\'{i}}fica (LNCC-SCAFMat) for the computer resources.

\bibliography{bib}


\end{document}